\documentclass[aps,prd,preprint,nofootinbib]{revtex4}
\usepackage{amsfonts}
\usepackage{mathrsfs}
\usepackage{graphicx}
\usepackage{amsmath}
\usepackage{amssymb}
\usepackage{epsfig}
\usepackage{graphicx}
\usepackage{slashed}
\usepackage{color}
\usepackage{multirow}
\usepackage{ulem}
\usepackage{adjustbox}

\usepackage{stmaryrd}
\newcommand{\bqa}{\begin{eqnarray}}
\newcommand{\eqa}{\end{eqnarray}}
\newcommand{\beq}{\begin{equation}}
\newcommand{\eeq}{\end{equation}}
\allowdisplaybreaks[1]

\graphicspath{{fig/}{dia/}} \DeclareGraphicsExtensions{.eps}
\begin{document}

\title{New extraction of CP violation in b-baryon decays}
\author{Chao-Qiang Geng, Xiang-Nan Jin, Chia-Wei Liu\footnote{chiaweiliu@ucas.ac.cn}, Zheng-Yi Wei and Jiabao Zhang}
\affiliation{	
	School of Fundamental Physics and Mathematical Sciences, Hangzhou Institute for Advanced Study, UCAS, Hangzhou 310024, China\\
	University of Chinese Academy of Sciences, 100190 Beijing, China
%	$^{3}$International Centre for Theoretical Physics Asia-Pacific (ICTP-AP), University of Chinese Academy of Sciences, Beijing 100190, China
}
\date{\today}
\vspace{0.6cm}

\begin{abstract}
\vspace{0.5cm}
We study CP violation in b-baryon decays  of $\Xi_b^- \to \Xi^-  D$  with $D = D^0, \bar{D}^0$ and  $D_i~(i=1,2)$. We find  that these baryonic decay processes provide  an ideal opportunity to measure the weak phase due to the absence of the relative strong phase. Explicitly, we relate $\bar{\rho}$ and $\bar{\eta}$ the CKM elements with the decay rate ratios of $R_i= \Gamma(\Xi_b^- \to \Xi^-D_{i}  ) / \Gamma( \Xi_b^- \to  \Xi^-D^0 )$ without  the charge conjugate states. As a complementary, we also examine the decay distributions of $\Lambda_b \to \Lambda(\to p \pi^- ) D$. There are in total 32 decay observables, which can be parameterized by 9 real parameters,  allowing the experiments to extract the angle $\gamma\equiv \arg(-V_{ud}V_{ub}^*/V_{cd}V_{cb}^*)$ in the CKM unitarity triangle. In addition, the feasibilities of the experimental measurements are discussed. We find that $\bar{\rho}$ and $\bar{\eta}$ can be extracted at LHCb Run3 from $\Xi_b^- \to \Xi^- D$, and a full analysis  of $\Lambda_b \to \Lambda(\to p \pi^-)D$ is available at LHCb Run4. 
\end{abstract}

\maketitle
%\newpage

\section{Introduction}
To complete the understanding of the standard model~(SM),
one of the important tasks  is to measure the Cabibbo-Kobayashi-Maskawa (CKM) quark mixing matrix elements. 
	So far, the experimental value of $\gamma\equiv \arg(-V_{ud}V_{ub}^*/V_{cd}V_{cb}^*)$ in the CKM unitarity triangle  comes exclusively from  the $ B $ meson decays~\cite{LHCb:2017hkl},  utilizing the $D^0 -\overline{D}^0$  mixing.  
The  simplest ways   are the  methods~\cite{Gronau:1990ra,Atwood:1996ci} of using  the $D$ meson two-body sequential decays for CP and flavor taggings.
However, their sensitivities are limited  by the smallness of the  two-body decay  branching ratios. 
To reduce the statistical uncertainties, one can analyze the Dalitz plot in the $D$ meson multibody decays  for the  two tagging methods~\cite{Grossman:2002aq}. 
Currently, the  most precise  value of $\gamma$ 
is  $\left(65.6_{-2.7}^{+0.9}\right)^{\circ}$ and $(65.8 \pm 2.2)^{\circ}$ from the CKMfitter~\cite{CKM} and UTfit~\cite{UT}, respectively.

On the other hand, the experimental interests on the b-baryon decays 
have been increasing rapidly. The evidences of  CP violation have been found in various multibody decays~\cite{LBCP}, while the decay width of $\Lambda_b \to \Lambda_c ^ + \tau^- \overline{\nu}_\tau$ has been measured for the first time~\cite{LHCb:2022piu}. 
Besides the branching ratios, the polarizations of the baryons provide fruitful observables in the experiments.  In addition, the forward-backward asymmetry of $\Lambda_b \to \Lambda \mu^+\mu^-$ has been studied at LHCb~\cite{Lambdamu}.
Notably, the polarization asymmetry of $\Lambda$ in $\Lambda_b \to \Lambda \gamma$ has been measured at LHCb for the first time~\cite{LHCb:2021byf}. 
Recently,  a complete analysis of $\Lambda_b \to \Lambda (\to p \pi^- )J/\psi(\to \mu^+\mu^-) $ has been performed~\cite{Jpsi}, where the polarization fraction
is found to be around $3\%$ at the centre-of-mass energy of $13$ TeV in $pp$ collisions.
 Despite these  progresses,  measurements on the decays associated  with a neutral $D$ meson 
are still lacking.

On the theoretical aspect,
the spin nature of the baryons is a double-edge sword, as 
it provides fruitful phenomenons~\cite{Observables,sizable,TV}, but   increases the complexity of the  quantitative studies. 
Most of the theoretical studies are performed by the factorization ansatz, within which the color-allowed decays can be estimated reliably~\cite{Theor,Zhu:2018jet,Geng:2020ofy}.
However, for the color-suppressed decays, one often has to introduce an effective color number by hand to explain the experiments. 
Fortunately,  with the helicity formalism,  we can analyze the kinematical  systems   without knowing the dynamical details~\cite{Gutsche:2013oea}.

The extraction of  $\gamma$ via $\Lambda_b \to \Lambda D$ with $D = D^0, \overline{D}^0, D_{1,2}$  has been given in  Refs.~\cite{Giri:2001ju, Zhang:2021sit}.
However, a systematic study of  the sequential decays is still missing.
In this work, we would use the helicity formalism to explore the sequential modes in the b-baryon decays.
%{\color{blue}
In contrast to the  orbital angular momentum analysis, the helicity formalism is perfectly compatible with  special relativity, and  it allows us to extract the information about  the baryon spin in a systematic way~\cite{Jpsi}. 
The difference between the orbital angular momentum and the helicity approaches is discussed  in Appendix A.  
%}

{
%\color{blue}
As mentioned, the extraction of $ \gamma $ comes exclusively from $ B $ meson decays,  and the motivations to extend it to baryon sectors are twofolds. On the theoretical aspect, it is  important  to  study  CP violation in baryons, since the matter-antimatter asymmetry in our universe is directly related to
 them, which can not be explained in the SM. 
On the experimental aspect, as  baryons carry a spin quantum number, it provides us  fruitful observables, allowing us to probe the SM further.
%, which will improve the sensitivity and feasibility of the extraction of the weak phase $ \gamma $.
For instance, the time-reversal violating observables can be constructed even for two-body baryonic decays. 
In our present work, we  concentrate on $ \Xi_b $  and $ \Lambda_b $ decays. As mentioned previously, a lot of observations of $ \Lambda_b $ have been made at LHCb,  and after the upgrade, there will also be enough $ \Xi_b $  to be created. This opens a new window to  reexamine $ \gamma $ obtained from the meson sectors. We propose to extract $ \gamma $ from baryon sectors, by the means of  measuring $ \Lambda_b\to\Lambda D,~\Xi_b\to\Xi D $ and their sequential decays. 
%}

This paper is organized as the follows. In Sec.~\ref{Formalism}, we present the formalism  related to  the possible physical observables.
In  Sec.~\ref{Numerical}, we show the numerical results based on
the factorization ansatz. We also explore  the experimental feasibilities for our results.
We conclude the study in Sec.~\ref{conclusion}.  

\section{Formalism}\label{Formalism}

%{\color{red}
We analyze the decays of ${\bf B}_b \to {\bf B}_n D$ with the helicity amplitudes
defined by 
\begin{equation}
H^\lambda _ j  \equiv \langle {\bf B}_n D_j , p \hat{z} , \lambda | {\cal H}_{eff} |{\bf B}_b, J_z = \lambda\rangle\,,
\end{equation}
where  ${\bf B}_{n(b)} = \Lambda_{(b)}$ and $ \Xi^{0,-}_{(b)} $ ,  $\lambda$ and $p\hat{z}$ are the helicity and 3-momentum of ${\bf B}_n$, respectively, $j=0,\overline{0},1,2$ denote  the $D$ mesons, and $J_z$ is the $z$ component of the angular momentum. The derivation and  physical meaning of the helicity amplitudes are sketched in Appendix A.
%}
In general, the positive helicity amplitudes  
have the ratios 
\begin{equation}\label{eq1}
H_0^+ : H_ { \overline{ 0}} ^+ : H_1^+ : H_2^+ =  \sqrt{2} : \sqrt{2} r^+ V: 1 +r^+V : 1 - r^+V\,,
\end{equation}
where 
  $r^+$ is defined by Eq.~\eqref{eq1} itself, and
  $V$ corresponds to the ratio of the CKM elements, given by
   $V = V_{ud}V_{ub}^*/ V_{cd} V_{cb}^*  = |V| e^{-i\gamma} $ with $V_{qq'}$ and $\gamma$ being the CKM elements and unitarity triangle, respectively.
The ratios of the negative  helicity ones can be obtained by substituting ``$-$'' for ''$+$'' in the superscripts.
The amplitudes are related to the CP conjugates as 
\begin{equation}\label{eq2}
H_0 ^+ =   -\overline{ H}_ { \overline{ 0}} ^- \,,~~~H_0 ^- =   -\overline{ H}_ { \overline{ 0}} ^+ \,,
\end{equation}
where we have taken $V_{cd}V_{cb}^*$ to be  real. 
The helicities flip signs due to the space inversion, and the minus signs are  attributed to the parity of the $D$ mesons. 

The amplitude ratios among the charge conjugates are 
\begin{equation}\label{eq3}
 \overline{ H}_0^- : \overline{ H}_ { \overline{ 0}} ^-  : \overline{ H}_1^- : \overline{ H} _2^- =   - \sqrt{2} r^+  V^* :   - \sqrt{2} :  - 1  - r^+ V^*  :  1 -  r^+ V^* \,,
\end{equation}
with  the positive  helicity ones given by interchanging ``$\pm $'' in the superscripts. 
Combining  Eqs.~(\ref{eq1})-(\ref{eq3}), the  16  complex  amplitudes are parameterized by one  real and four complex parameters, given by 
\begin{equation}\label{parameters}
|H^+_0| \,,~~\tilde{H} = \frac{ H^-_0}{H_0^+}\,,~~~r^\pm \,,~~~V\,,
\end{equation}
which remarkably 
simplify the analysis.

The decay widths for ${\bf B}_b \to {\bf B}_n D_j$   are   given as 
\begin{equation}
\Gamma_j = \frac{|\vec{p}| }{16 \pi M_{{\bf B} _b} } \left(
\left| H_j^+ \right|^2  + \left| H_j^- \right|^2 
\right)\,,
\end{equation}
where  $\vec{p}$ is the 3-momentum of the daughter particle, and  $M_{{\bf B}_b}$ denotes the mass of ${\bf B}_b$.  
%{ \color{blue}
The full sequential decays ${\bf B}_b\to {\bf B}_ n (\to {\bf B}_n' \pi)  D_j$, where ${\bf B}_n' = \Lambda(p)$ for ${\bf B}_b = \Xi_b (\Lambda_b)$,  offer three additional observables in the angular distributions.  
The derivation is given  in Appendix A, and the result is sorted as follows:
%} 
\begin{align}\label{level2}
	&\begin{aligned}
{\cal D}_j  ( \vec{\Omega}) = 	&\frac{1}{\Gamma_j}\frac{\partial^3 \Gamma_j }{\partial \cos \theta \partial \cos\theta_1 \partial \phi }=\frac{1}{8\pi} [1 +P_b \alpha_n \cos \theta \cos \theta_1   +\\
		&\qquad \alpha_j (\alpha_n \cos \theta_1 + P_b \cos \theta)+P_b \alpha_n(\beta_j \sin \phi 
		-\gamma_j \cos \phi 
		)  \sin \theta \sin \theta_1],
	\end{aligned}
\end{align}
where $P_b$ is the polarized fraction of ${\bf B}_b$, depending on its production, and 
the definitions of the angles are shown in FIG.~\ref{Quark} with $( \theta, \phi)$ and  $\theta_1$   determined at the rest frames of ${\bf B_b}$ and ${\bf B}_n$, respectively. In Eq.~\eqref{level2}, $\alpha_n$ is the up-down asymmetry parameter of ${\bf B}_n \to {\bf B}_n' \pi$, and  $\alpha_j$, $\beta_j$ and $\gamma_j$  are given as 
\begin{equation}\label{eq7}
	\alpha_j= \frac{1  -|\tilde{H}_j|^2    }{1  + |\tilde{H}_j|^2  }\,,~~~\beta_j = \frac{- 2 \text{Im}(\tilde{H}_j )}{1  + |\tilde{H}_j|^2 }\,,~~~\gamma _j= \frac{2 \text{Re}(\tilde{H}_j )}{1  + |\tilde{H}_j|^2 }\,,
\end{equation}
respectively, where
$\tilde{H}_j$ are   defined by 
\begin{equation}
	\tilde{H}_j = H_j^- / H_j^+ \,,
\end{equation}
with the explicit parametrizations  in Table~\ref{table3}\,,
$\alpha_j$ describe the polarized asymmetries of the daughter baryons, and $\beta_j$ represent T violation for the absence of strong phases~\cite{Geng:2020ofy}. 

%{\color{blue}
To measure  $\alpha_j$, $\beta_j $ and $\gamma_j $, it is convention to recast them in the forms based on   the numbers of events, $N$, given as 
\begin{eqnarray}
&&\alpha_j = 
\frac{2}{\alpha_n }
\frac{
N(\hat{p}_{{\bf B}_n'} \cdot \hat{p}_{D_j} < 0)  - N(\hat{p}_{{\bf B}_n'} \cdot \hat{p}_{D_j} > 0)
}{
N(\hat{p}_{{\bf B}_n'} \cdot \hat{p}_{D_j}>0) + N(\hat{p}_{{\bf B}_n'} \cdot \hat{p}_{D_j}<0)
}\,, \nonumber\\
&&\beta_j = 
\frac{8}{P_b \alpha _n \pi }
\frac{
	N\left( (\hat{p}_{{\bf B}_n'}   \times \hat{p} _{D_j} ) \cdot \hat{n}_{{\bf B}_b }  >  0\right)  - 	N\left( (\hat{p}_{{\bf B}_n'}   \times \hat{p} _{D_j} ) \cdot \hat{n}_{{\bf B}_b }  <  0\right) 
}{
	N\left( (\hat{p}_{{\bf B}_n'}   \times \hat{p} _{D_j} ) \cdot \hat{n}_{{\bf B}_b }  >  0\right)  +	N\left( (\hat{p}_{{\bf B}_n'}   \times \hat{p} _{D_j} ) \cdot \hat{n}_{{\bf B}_b }  <  0\right) 
}\,, \\
&&\gamma_j   = 
\frac{8}{P_b \alpha _n  \pi }
\frac{
	N\left( (\hat{p} _{D_j}\times \hat{n}_{{\bf B}_b} )   \cdot (\hat{p} _{D_j}\times \hat{p} _{{\bf B}_n'}) > 0\right)  - 			N\left( (\hat{p} _{D_j}\times \hat{n}_{{\bf B}_b} )   \cdot (\hat{p} _{D_j}\times \hat{p} _{{\bf B}_n'}) < 0\right)  
}{
	N\left( (\hat{p} _{D_j}\times \hat{n}_{{\bf B}_b} )   \cdot (\hat{p} _{D_j}\times \hat{p} _{{\bf B}_n'}) > 0\right)   +	N\left( (\hat{p} _{D_j}\times \hat{n}_{{\bf B}_b} )   \cdot (\hat{p} _{D_j}\times \hat{p} _{{\bf B}_n'}) < 0\right)
}   \,,\nonumber
\end{eqnarray}
respectively, where the equations hold at the limit of $N\to \infty$. 
%}

 \begin{table}
	\caption{ The parameterization of $\tilde{H}_j$.   }
	\label{table3}
	\begin{tabular}{lcc}
		\hline
		\hline
		$D $ & ~$\tilde{H}_j({{\bf B}_b \to {\bf B}_n D})$ ~&~$\tilde{H}_j({\overline {\bf B}}_b \to {\overline {\bf B}}_n D) $~ \\
		\hline
		$ D^0 $&$ \tilde{H}  $&$ \frac{\displaystyle r^+}{\displaystyle r^-} \tilde{H}^{-1} $\\
		$ \overline{D}^{ 0} $&$ \frac{\displaystyle r^-}{\displaystyle r^+}\tilde{H} $&$\tilde{H}^{-1} $\\
		$ D_1 $&$  \frac{\displaystyle 1+r^+V}{\displaystyle 1+r^-V}\tilde{H}  $&$  \frac{\displaystyle 1+r^+V^*}{\displaystyle 1+r^-V^*} \tilde{H}^{-1} $\\
		$ D_2 $&$  \frac{\displaystyle 1-r^+V}{\displaystyle 1-r^-V}\tilde{H}  $&$  \frac{\displaystyle 1-r^+V^*}{\displaystyle 1-r^-V^*} \tilde{H}^{-1} $\\
		\hline
		\hline
		
	\end{tabular}
\end{table}

The CP violating asymmetries are constructed as
\begin{equation}\label{CPv}
	A_j^{CP} = \frac{\Gamma _ j - \overline{ \Gamma }_{\overline{j}  } }{\Gamma_j + \overline{ \Gamma }_{\overline{j}  }}  \,,~~~ \Delta \xi_j =  \frac{1}{2}\left(
	\xi_j  + \overline{\xi} _{\overline{j} }   
	\right),
	~~~\Delta \gamma _j=\frac{1}{2}\left(
	\gamma_j - \overline{\gamma}_{\overline{j}} 
	\right) \,,~~~ \text{with}~\xi_j = \alpha_j, \beta_j,
\end{equation}
where 
the overlines on $\Gamma_j$, $\xi_j$ and $\gamma_j$  correspond to  the charge conjugate ones of  the baryons, and
$\overline{j}= \overline{0}, 0 , 1, 2 $ for $j = 0, \overline{0}, 1, 2 $\,, respectively. Note that   
 $A^{CP}_j$ are the  direct CP asymmetries, 
and  the others are the CP violating obeservables in the decay angular distributions.

\begin{figure}[t]
	\includegraphics[width=.48\textwidth]{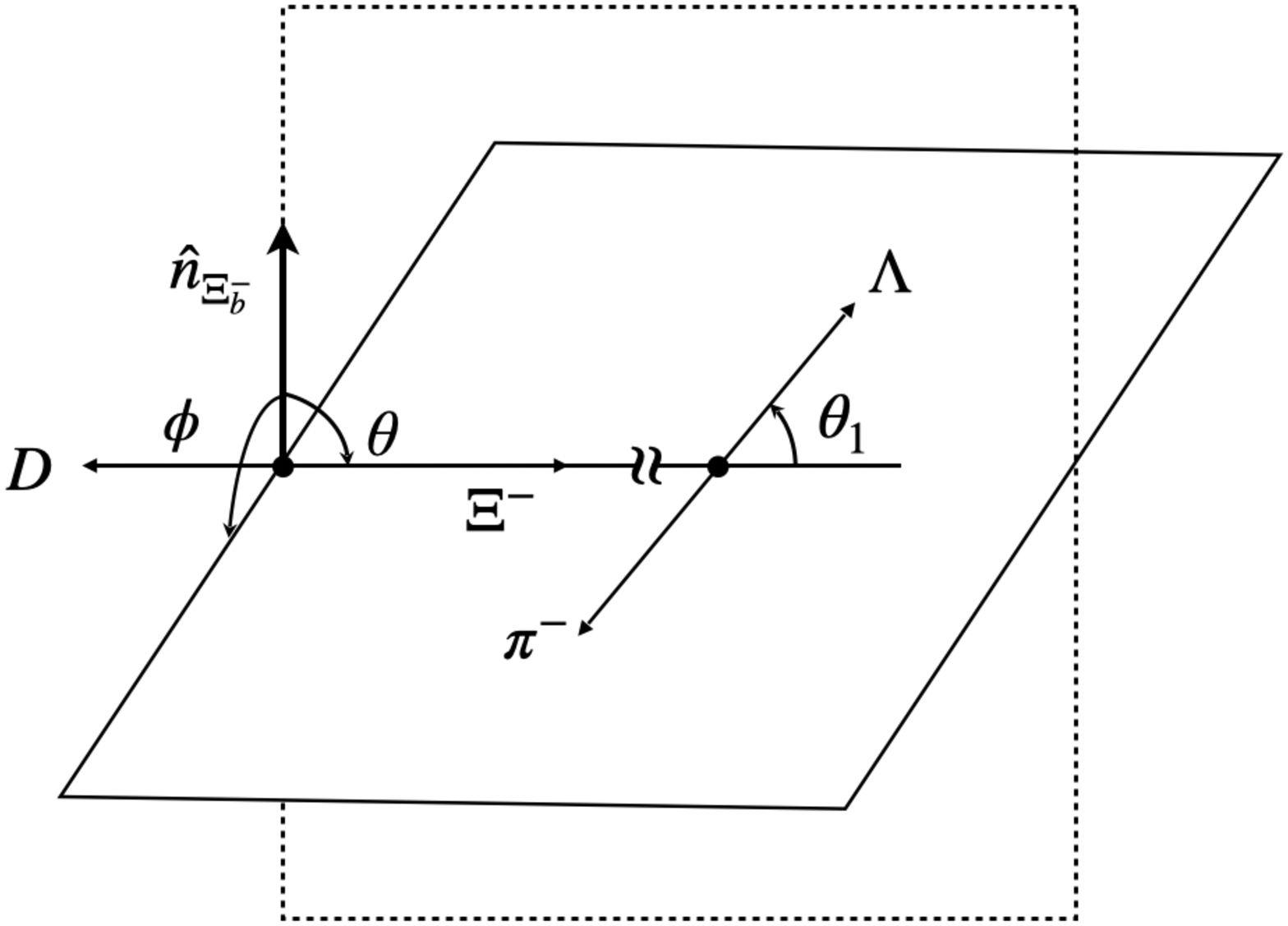}  
		\includegraphics[width=.48\textwidth]{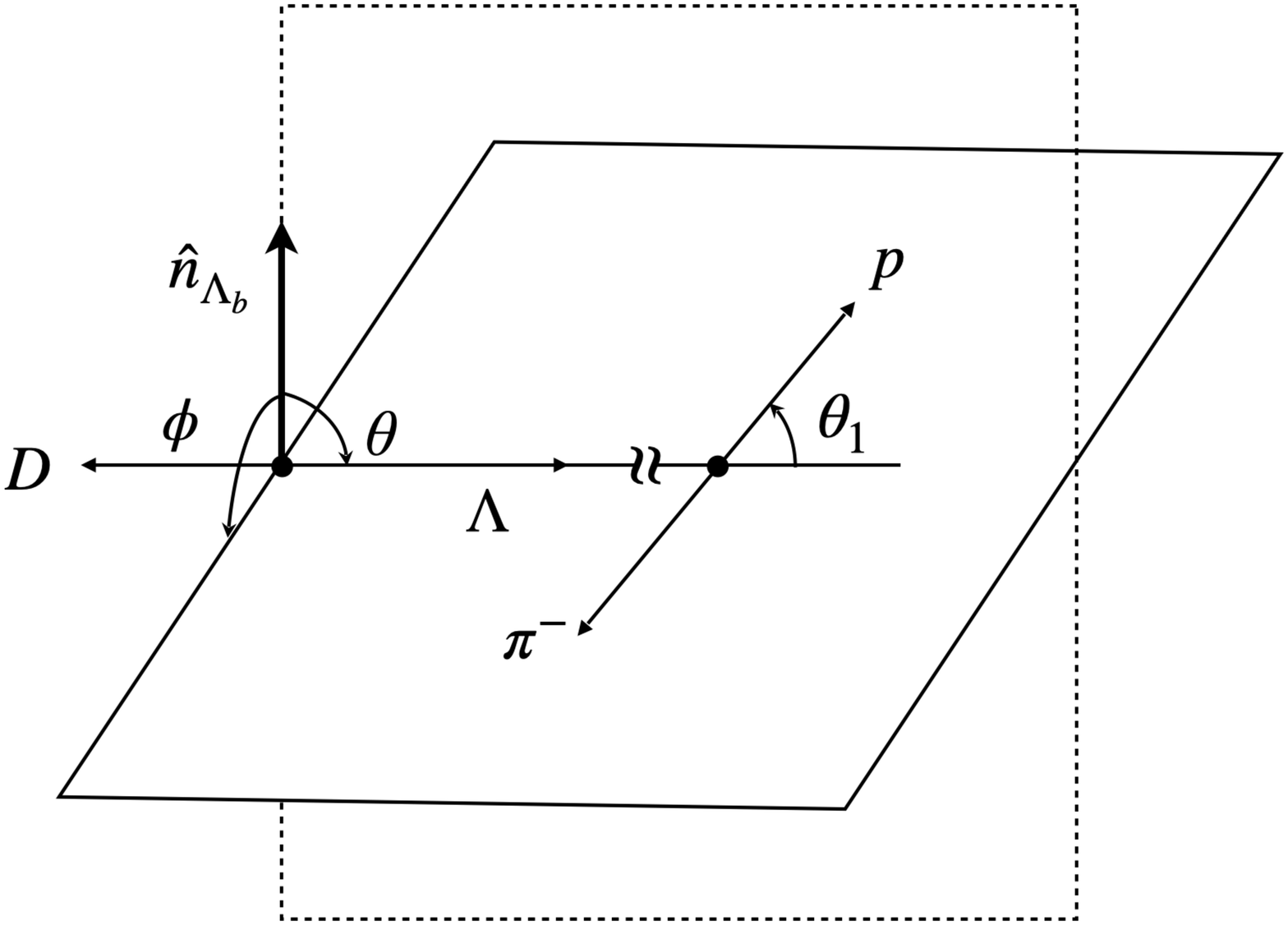}  
	\caption{  Decay planes for ${\bf B}_b \to {\bf B}_n(\to {\bf B}_n' \pi)  D$. } 
	\label{Quark}
\end{figure}

To get a clearer view of $\Delta \xi _j $ as well as $\Delta \gamma_j$,  we   rewrite
the decay parameters  as
\begin{eqnarray}\label{eq10}
	&& \xi_j ^{CP} = \Delta \xi _ j + \frac{1}{2}A_j^{CP} \left(
	\xi  _ j -  \overline{\xi} _{\overline{j}  }
	\right)\,, ~~%= \frac{ {\lambda \Gamma  + \overline{\lambda} \overline{\Gamma}  } } { \Gamma + \overline{ \Gamma } } - \frac{1}{2}A_{CP} \left(\lambda  -  \overline{\lambda} 	\right)\,,\nonumber\\
	~	\gamma_j^{CP}  = \Delta \gamma _ j + \frac{1}{2}A_j ^ {CP} \left(
	\gamma_ j  + \overline{\gamma} _{\overline{j}  }
	\right)\,,% \frac{ {\gamma \Gamma  -\overline{\gamma} \overline{\Gamma}  } } { \Gamma + \overline{ \Gamma } } - \frac{1}{2}A_{CP} \left(\gamma  + \overline{\gamma} 	\right)\,.
\end{eqnarray}
with 
\begin{equation}
	\xi_j^{CP} =  \frac{ {\xi_j \Gamma_j   + \overline{\xi}_{\overline{j}  } \overline{\Gamma} _{\overline{j}  } } } { \Gamma _j + \overline{ \Gamma }_{\overline{j}  } } \,,~~~\gamma_j^{CP}=  \frac{ {\gamma_j \Gamma _j -\overline{\gamma}_{\overline{j}  } \overline{\Gamma} _{\overline{j}  } } } { \Gamma _j + \overline{ \Gamma } _{\overline{j}  }} \,.
\end{equation}
As there is only one weak phase in the decay channels with $D^0$ and $\overline{ D}^0$, we have 
\begin{equation}\label{eq11}
A_0^{CP} = A_{\overline{0}}^{CP} =  \xi^{CP}_0 =  \xi^{CP}_{\overline{ 0}} =  
 \gamma^{CP}_0 =  \gamma_{\overline{ 0}} ^{CP}
=  0 \,,
\end{equation}
 derived from Eqs.~(\ref{eq1})-(\ref{eq3})\,.
The right sides of the two equations in Eq.~\eqref{eq10} can be measured from the experiments, whereas the left ones can be written down in  compact ways as 
\begin{equation}\label{eq12}
A_{1,2}^{CP} = \frac{\pm 2 {\cal X}_+}{\langle H_{1,2}^2 \rangle}\,,~~
\alpha_{1,2}^{CP} = \frac{\pm  2 {\cal X}_-}{\langle H_{1,2}^2 \rangle}\,,
~~\beta_{1,2}^{CP} =  \frac{\pm 2 \text{Re}({\cal Y} ) } {\langle H_{1,2}^2 \rangle}\,,~~\gamma_{1,2}^{CP} =  \frac{\pm  2 \text{Im} ({\cal Y} ) } {\langle H_{1,2}^2 \rangle}\,,
\end{equation}
where 
\begin{eqnarray}
&&{\cal X}_\pm = - \text{Im}(V) \text{Im}\left(
r^+  \pm  |\tilde{H}|^2 r^-
\right)\,,~~{\cal Y } = \text{Im} (V) \tilde{H }_j \left(
r^{+*} - r^-
\right)\,,\nonumber\\
&& \langle H_{1,2}  ^2 \rangle = 1 + |\tilde{H}|^2 \pm 2 \text{Re}(r^+ + r^-|\tilde{H}|^2) \text{Re}(V) + \left(|r^{+}|^2+|\tilde{H}r^{- }|^2 \right)|V|^2\,.
\end{eqnarray}
It is then straightforward to see that the observables defined in Eq.~\eqref{CPv} are CP odd, as they are proportional to Im$(V)$.

It is not a coincidence
that the CP violating asymmetries of $D_1$ and $D_2$ differ minus signs in the numerators of Eq.~\eqref{eq12}, as  can be seen from  the following identities:
\begin{eqnarray}\label{eq15}
	&&  ( \Gamma_1  + \overline{\Gamma}_1  ) \xi^{CP} _ 1 +   ( \Gamma_2  + \overline{\Gamma}_2 )  \xi^{CP} _2 =    ( \Gamma_0  + \overline{\Gamma}_{\overline{0} } )  \xi ^{CP}_ 0 +   ( \Gamma_{\overline{0} }  + \overline{\Gamma}_0 )   \xi ^{CP}_{\overline{0}}  = 0\,,\nonumber\\
		&& ( \Gamma_1  + \overline{\Gamma}_1  )  \gamma^{CP} _ 1 +     ( \Gamma_2  + \overline{\Gamma}_2 )\gamma ^{CP} _2 =    ( \Gamma_0  + \overline{\Gamma}_{\overline{0} } )  \gamma ^{CP}_ 0 +   ( \Gamma_{\overline{0} }  + \overline{\Gamma}_0 )   \gamma ^{CP}_{\overline{0}} = 0\,.
\end{eqnarray}
In Eq.~\eqref{eq15},
the first equality  comes from that the physical quantities are independent of  the basis (either flavor or CP), and  the second one is due to Eq.~\eqref{eq11}.

\section{Numerical Results}\label{Numerical}
To analyze the decays quantitatively, we begin with  the  effective Hamiltonian~\cite{Buras:1991jm}, given by 
\begin{eqnarray}
&&{\cal H}_{eff} = \frac{G_F}{\sqrt{2}}
\left[
V_{cb}V_{us}^* \left(
C_1O^c_1 + C_2 O^c_2  \right)  + V_{ub}V_{cs}^* \left(
C_1O^u_1 + C_2 O^u_2  \right)  
\right] +h.c.\,,
\end{eqnarray}
with 
\begin{eqnarray}
&&O_1^c = (\overline{ c}_\beta  u_\alpha  )_{V-A}  (\overline{ d }_\alpha b_\beta )_{V-A} \,,~~~~O_2^c = (\overline{ c }_\alpha  u_\alpha  )_{V-A}  (\overline{ d}_\beta b_\beta )_{V-A} \,,\nonumber\\
&&O_1^u = (\overline{ u}_\beta  c_\alpha  )_{V-A}  (\overline{ d }_\alpha b_\beta )_{V-A} \,,~~~~O_2^u  = (\overline{ u }_\alpha  c_\alpha  )_{V-A}  (\overline{ d}_\beta b_\beta )_{V-A} \,,
\end{eqnarray}
where $G_F$ is the Fermi constant, $C_{1,2}$ are  the Wilson coefficients,  $\alpha$ and $\beta$ correspond to  the color indices, and $h.c.$  represents the Hermitian conjugate.
Note that we have used the Fierz transformation to sort the operators.

%\textcolor{red}{
The quark diagrams of ${\bf B}_b \to {\bf B}_n D$  are shown in FIG.~\ref{non}. 
There is only one possible type of quark diagrams for $\Xi_b^-\to \Xi^-D$ 
%({\color{red}
(the left one in FIG.~2).
%}.
In contrast, the decays of  $\Lambda_b$  have two extra nonfactorizable  diagrams (the middle and  right ones in FIG.~2), which would introduce different strong phases and increase the complexity of the analysis. 
In the following, unless stated otherwise, we  concentrate on $\Xi_b^- \to \Xi^- D$. 
%}

\begin{figure}[t]
	\includegraphics[width=.3\textwidth]{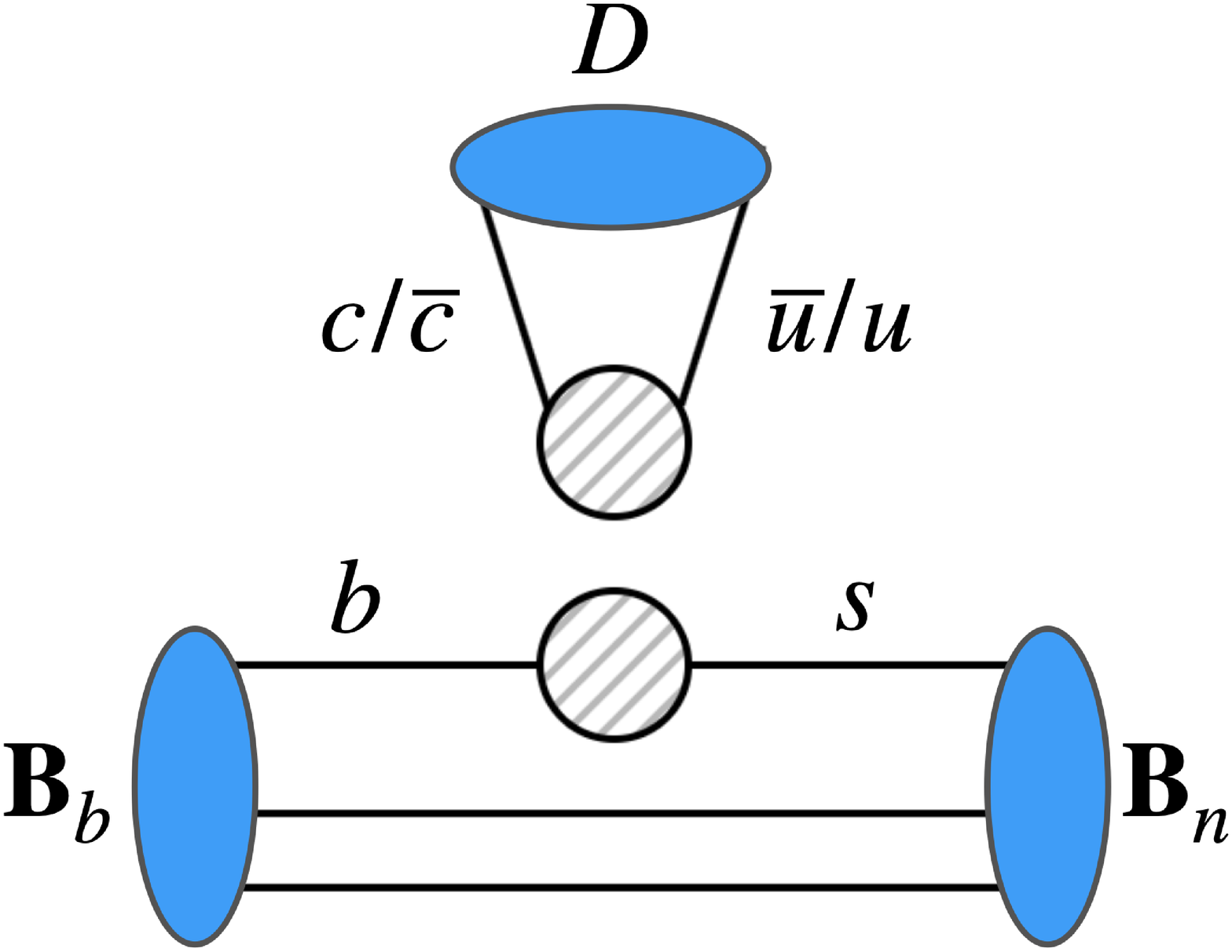}  
	\includegraphics[width=.3\textwidth]{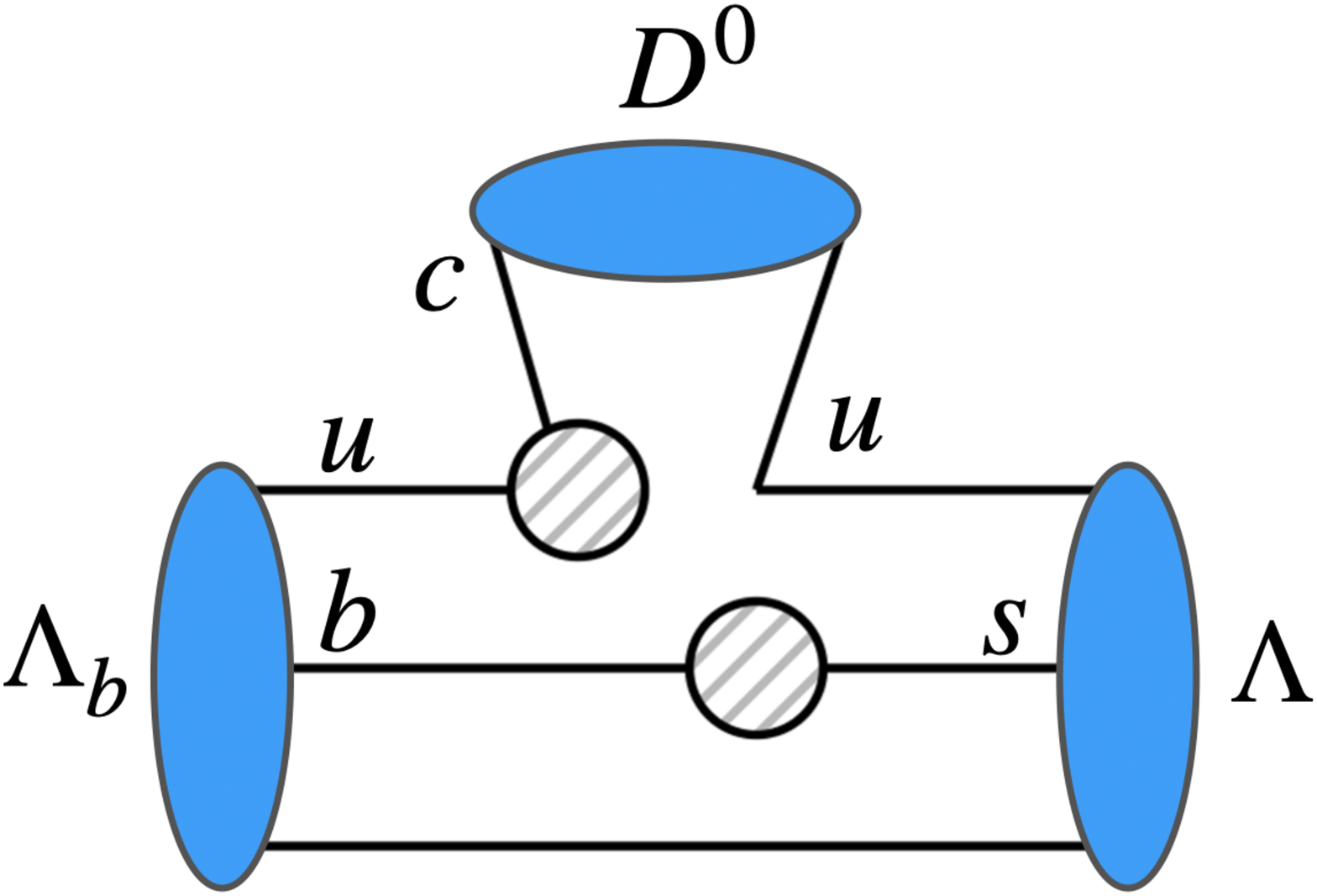}  
	\includegraphics[width=.3 \textwidth]{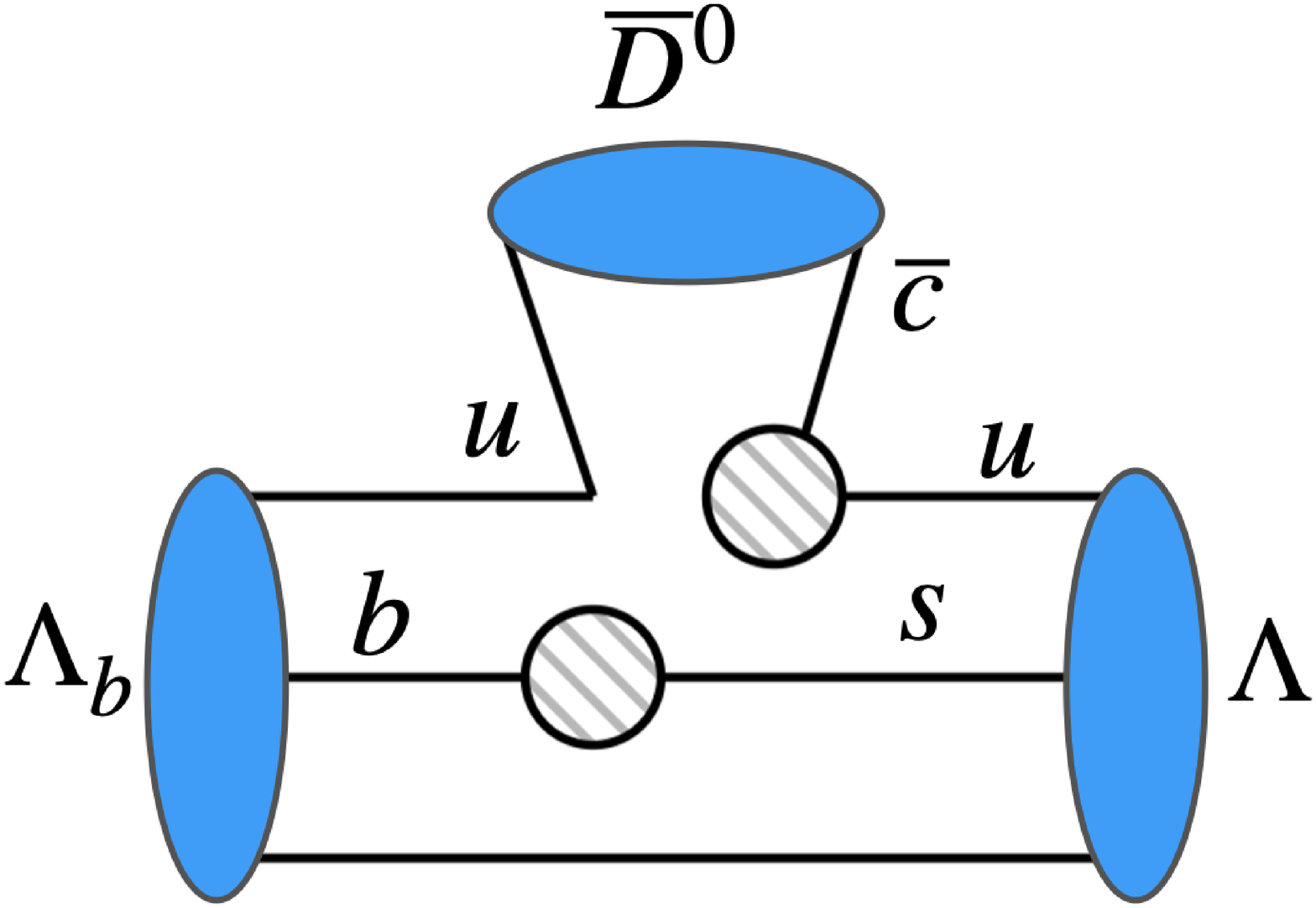}  
	\caption{  Quark diagrams of the b-baryons decays%. The  diagram at the left  is factorizable, whereas the others are not.
	}
	\label{non}
\end{figure}

The amplitude ratios of  $\Xi_b^- \to  \Xi^- D$ can be naively read off in a model independent way from FIG.~\ref{Quark}. 
%\textcolor{red}{
As  $\Xi_b^- \to \Xi^- D^0 $ and  $\Xi_b^- \to \Xi^- \overline{D}^0 $
share the same diagram, they 
 receive the same corrections from the strong interactions. Thus, we  deduce that  $r^\pm=1$ in Eq.~\eqref{eq1} for $\Xi_b^- \to \Xi^-D$, leading to 
%}
\begin{eqnarray}\label{ratios}
\Gamma_0 :\Gamma_{\overline{0} }: \Gamma_1 :\Gamma_2  
&=&  2 :2\overline{\rho}^{ 2} + 2 \overline{\eta}^{ 2} : 1 + 2\overline{\rho} + \overline{\rho}^{ 2} +\overline{\eta}^{ 2}:1 -  2\overline{\rho} + \overline{\rho}^{ 2} +\overline{\eta}^{ 2}\,,
\end{eqnarray}
to ${\cal O}(\lambda^4) $ precision with $\lambda$, $\overline{\rho}$ and $\overline{\eta}$ the Wolfenstein parameters\footnote{Here, $\lambda=|V_{us}|/\sqrt{|V_{ud}|^2+|V_{us}|^2}$ and $\bar{\rho}+i \bar{\eta}=-(V_{ud}V_{ub}^*)/(V_{cd}V_{cb}^*)$.} for the CKM matrix~\cite{CKM}. 
As the total branching ratios are independent of the  basis~(flavor or CP), we have 
\begin{equation}
\Gamma_{ 0 } + \Gamma_{ \overline{0} } = \Gamma_1 + \Gamma_2\,,
\end{equation}
which can also be easily seen from  Eq.~\eqref{ratios}\,. 
Hence, there have only  two  independent ratios along with the two parameters $(\overline{\rho}, \overline{\eta})$. Clearly,  it is possible to extract both $\overline{\rho}$ and $\overline{\eta}$ from the experiments, 
given by 
\begin{equation}\label{heart}
\overline{\rho} = \frac{1}{2}\left(
R_1-R_2
\right)\,,~~~ \overline{\eta} = \sqrt{ 
R_1+R_2  - \frac{1}{4} \left( R_1  - R_2 
\right)^2
-1}\,,
\end{equation}
with  $R_{1,2} = \Gamma_{1,2}/\Gamma_0$\,. 
Remarkably, the extractions do {\it not} require  the  charge conjugate states.
%\textcolor{red}{
We emphasize that Eqs.~\eqref{ratios} and \eqref{heart} are model independent  based on the  analysis from the quark diagrams. 
%}

To estimate the results in the experiments, we adopt the framework of the  na\"ive factorization. The amplitudes are  then read as 
\begin{equation}
 \frac{G_F}{\sqrt{2}} a_2  V_{cb} V_{us} ^*  \langle D^0 |  
    (\overline{c} u )_{V-A}  
| 0 \rangle \langle \Xi^- | (\overline{s} b)_{V-A}| \Xi^-_b\rangle\,,
\end{equation}
where $a_2$ is the effective Wilson coefficient for the color-suppressed decays. For the numerical results, we take the large $N_c$ limit leading to $a_2 = c_2= -0.365$~\cite{Buras:1991jm}. 
The baryon transition matrix elements can be further decomposed as
\begin{eqnarray}
	\langle \Xi^- | \overline{s} \gamma^\mu b | \Xi^-_b \rangle 
	&=&\overline{u} \left(
	f_1(M_D^2) \gamma^\mu - f_2 (M_D^2)i \sigma_{\mu \nu} \frac{p_D^\nu}{ M_{\Xi^-_b}}   +f_3(M_D^2) \frac{p_D^\mu}{M_{\Xi^-_b}}
	\right)u_b\,,\nonumber\\
	\langle \Xi^- | \overline{s} \gamma^\mu \gamma^5 b | \Xi^-_b\rangle 
	&=&\overline{u} \left(
	g_1(M_D^2) \gamma^\mu - g_2 (M_D^2)i \sigma_{\mu \nu} \frac{p_D^\nu}{ M_{\Xi^-_b}}   +g_3(M_D^2) \frac{p_D^\mu}{M_{\Xi^-_b}}
	\right)\gamma^5 u_{b}\,,
\end{eqnarray}
where $u_{(b)}$ is the Dirac spinor of $\Xi_{(b)}^- $, 
$M_{D(\Xi_b^-)} $ and $p^\mu_D$  are the masses and the 4-momentum of $D(\Xi_b^-)$, respectively, and $f_i$ and $g_i$  are the form factors with $i = 1,2$ and $3$. 
The helicity amplitudes are related to the form factors as 
\begin{equation}
	H^\pm_0 = Q_+ A \mp Q_- B\,,
\end{equation}
where 
\begin{eqnarray}
&&Q_\pm = \sqrt{(M_{\Xi^-_b} \pm M_{\Xi^-})^2 - M_D^2}\nonumber\\
&&	A =  \frac{G_F}{\sqrt{2}}a_2f_DV_{cb} V_{us}^* 
	\left[
\left(M_{\Xi^-_b}  -M_{\Xi^-}\right)   	f_1 + \frac{M_{D}^2}{M_{\Xi^-_b }   }f_3 
	\right]\,,\nonumber\\
&& B =  \frac{G_F}{\sqrt{2}}a_2 f_DV_{cb} V_{us}^* 
\left[
 \left( M_{\Xi^-_b}  + M_{\Xi^-}\right)   	g_1 - \frac{M_{D}^2}{M_{\Xi^-_b }   }g_3 
\right]\,.
\end{eqnarray}
The rest of the amplitudes can be obtained  by taking
$r^\pm  =1 $ in Eq.~\eqref{eq1}\,.

 In this work, 
the form factors are calculated by   the homogeneous bag model~\cite{Geng:2020ofy}, in which the center motions of the hadrons are removed by the linear superposition of infinite bags, allowing the form factors to be calculated consistently.  
Particularly, with the homogeneous bag model, the experimental branching ratios of $\Lambda_b\to \Lambda_c^+ \pi^+/K^+$ and  $\Lambda_b\to p \pi^+/K^+$ can be well explained~\cite{Geng:2020ofy,sizable}.
All the  model parameters can be  extracted from the mass spectra, given as ~\cite{Zhang:2021yul}
\begin{equation}
R = (4.6\pm 0.2 ) ~\text{GeV}^{-1}\,,~~~M_{u,d} =0\,,~~~M_s = 0.28~\text{GeV}\,,~~~M_b = 5.093~\text{GeV}\,,
\end{equation}
where $R$ is the bag radius.
For  the detail derivations, the readers are referred to Ref.~\cite{Geng:2020ofy}\,.

In Table~\ref{table}, we list our numerical of the decay widths and observables.
At the chiral limit,  the emitted $s$ quark due to  the weak interaction is essentially left-handed, leading to $\alpha_j\approx-1$. As a result,  we  have that  $\beta_j =0 $ for  $\tilde{H}$ being  real within the factorization framework. In addition,  as there is no relative strong phase, the CP violating effects are absent.

\begin{table} 
	\caption{Decay widths and  observables
	}
	\label{table}
	%\scriptsize
	\setlength{\tabcolsep}{2mm}{
		%	\begin{center}
			\begin{tabular}[t]{llccccc}
				\hline
				Baryon&		$D$ & $\Gamma/\Gamma_0$  & $10^6{\cal B}$&$\alpha_j$  &$\gamma'_j$\\
				\hline
				\multirow{4}{*}{$\Xi_b^- \to \Xi^-$}&$D^0$    &  $\equiv 1 $   &  $9.7\pm 1.6  $ &\multirow{4}{*}{$-0.99\pm 0.01  $} &\multirow{4}{*}{$0.06 \pm 0.02$}\\
				&$D_1$&$0.71\pm 0.02 $&$6.9\pm 1.2 $\\
&$D_2 $&$0.43\pm 0.01 $&$4.2\pm 0.7$\\

				&$\overline{ D^0}$&$0.14\pm 0.01$ & $1.4\pm 0.3 $ \\
				\hline
				\multirow{4}{*}{$\Lambda_b \to \Lambda$}&$D^0$    &  $\equiv 1 $   &  $6.6 \pm 0.6  $ &\multirow{4}{*}{$-0.99\pm 0.01  $} &\multirow{4}{*}{$0.06 \pm 0.02$}\\
				&$D_1$&$0.71\pm 0.02 $&$4.7\pm 0.5 $\\
&$D_2 $&$0.43\pm 0.01 $&$2.9\pm 0.3$\\
				&$\overline{ D^0}$&$0.14\pm 0.01$ & $0.9\pm 0.1 $ \\
				\hline
		\end{tabular}}
	\end{table}

The results of $\Lambda_b \to \Lambda D $, estimated with the na\"ive factorization, are also given  to compare with  those in  the literature. 
Our  prediction of ${\cal B}(\Lambda_b \to D^0 \Lambda) $  is roughly 1.2 times larger than the one in Ref.~\cite{Giri:2001ju} and twice larger than that in Ref.~\cite{Zhu:2018jet}.  It is attributed to the use of  a larger $a_2$ in our study.  Since $\alpha_j$ are independent of $a_2$,  the predicted values of $\alpha_j$ are well consistent  with those in  Ref.~\cite{Zhu:2018jet}, which are direct consequences from the factorization approach.

 %The decay channels we proposed have not been measured yet. 

 \begin{figure}[t]
 	\includegraphics[width=.7\textwidth]{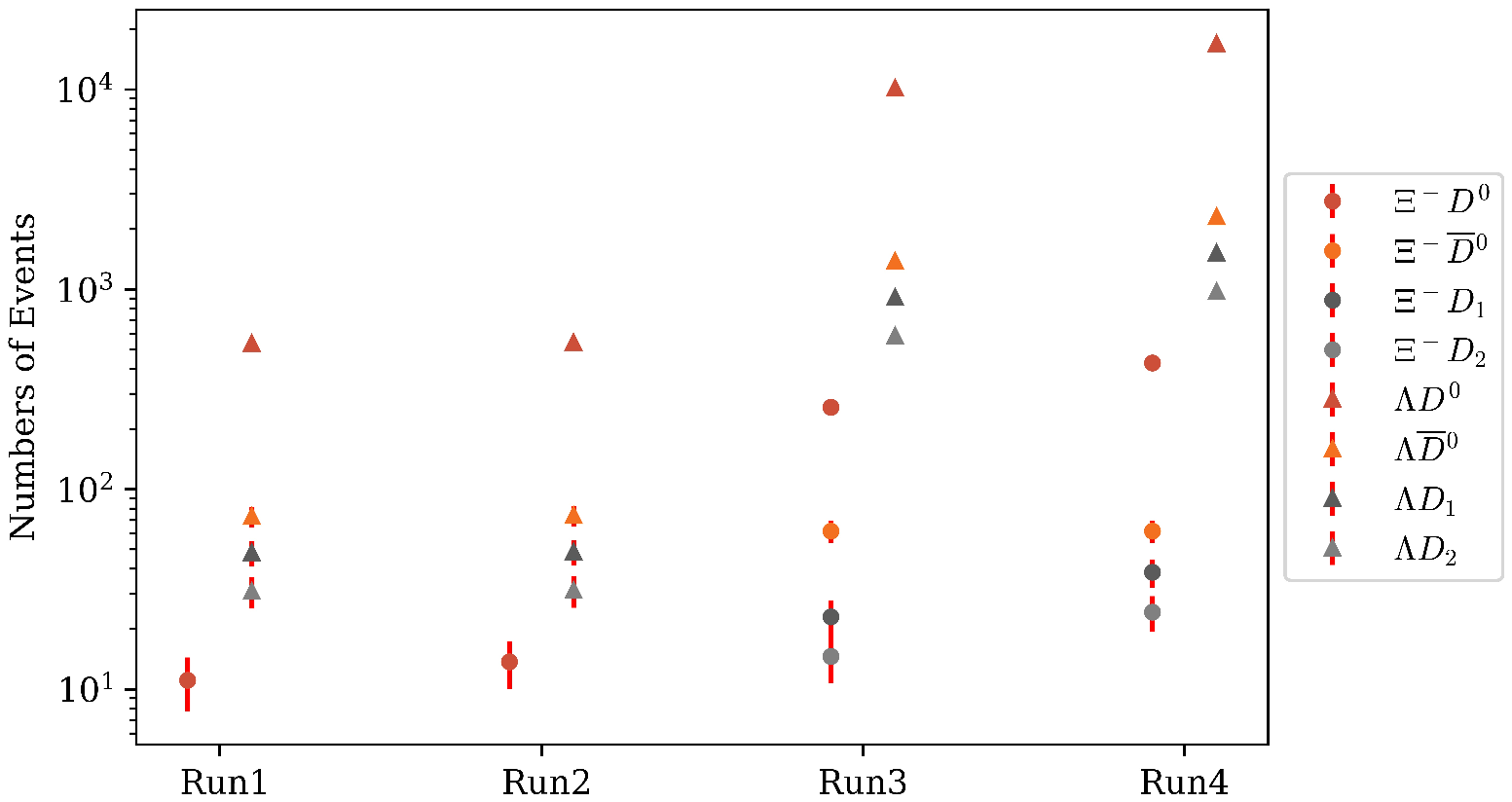}  
 	\caption{  The estimated values of $N(  \Lambda_b \to \Lambda (\to p \pi^-)  D)$ and    $N(\Xi _b ^ - \to  \Xi^- ( \to \Lambda \pi) D )$ at LHCb, where the red ones represent  the statistical uncertainties. } 
 	\label{Baryon}
 \end{figure}

The possible sequential decays for the flavor and CP taggings are~\cite{BESIII:2020khq}
\begin{equation}
\begin{array}{l} 
D^0: K^- \pi^+\,,~K^- \pi^+\pi^0\,,~K^-\pi^+\pi^+\pi^-\,,~K^-e^+\nu_e\,,\\
D_1:K_S^0\pi^0\,,K_S^0\eta\,,K_S^0\omega\,,K_S^0\eta'\,,\\
D_2:K^+K^-\,,\pi^+\pi^-\,,K_S^0\pi^0\pi^0\,,K_L^0\pi^0\,,\pi^+\pi^-\pi^0\,.\\
%\overline{D}^0: K^+ \pi^-\,,~K^+ \pi^-\pi^0\,,~K^+\pi^+\pi^-\pi^-\,,~K^+e^-\nu_e\,.
\end{array}
\end{equation}
By crunching up their branching ratios, the ideal 
  efficiencies are  $30\%$ 
 for the flavor tagging and 
 $(3.8, 4.0) \%$ for  $D_{1,2}$. 
 Accordingly,
 we give  estimations in FIG.~\ref{Baryon} through 
  \begin{eqnarray}
 	\label{Num events}
 	&&N(\Lambda_b \to \Lambda (\to p \pi^-)  D) = {\cal N}_{\Lambda_b}\times\mathcal{B}(\Lambda_b\to\Lambda(\to p\pi^-)D)\times\epsilon_{\text{Dtag}}\times\epsilon_{\text{exp}},\nonumber\\
 		&&N(\Xi_b^- \to \Xi^- (\to \Lambda \pi^-)  D) = {\cal N}_{\Xi_b^-}\times\mathcal{B}(\Xi_b\to\Xi^-(\to \Lambda\pi^-)D)\times\epsilon_{\text{Dtag}}\times\epsilon_{\text{exp}},
 \end{eqnarray} 
where $N$ represent the numbers of the observed events,
while $ {\cal N}_{{\bf  B}_b} $ are the produced numbers of ${\bf  B}_b=\Lambda_b,\Xi_b^-$, at the experiments.
The estimated values of ${\cal N}_{{\bf B}_b}$ can be found in  Appendix B. Here, 
 $\epsilon_{\text{exp}}$ are taken to be $7.25\%$ and $0.96\%$ for $\Lambda_b\to \Lambda(\to p \pi^-) D$ and $\Xi_b\to \Xi(\to p \pi^-) D$~\cite{LHCb:2019sxa}, respectively.

From the figure, we have  that $N(\Xi_b^- \to \Xi^-(\to \Lambda \pi^-) D^0,\overline{ D}^0 , D_1 ) = (200,  50, 20) $ at LHCb Run3, which are  sufficient for measuring $\overline{\eta}$ and $\overline{\rho}$ via Eq.~\eqref{heart}.  At LHC Run4, $N(\Lambda_b \to \Lambda(\to p \pi^-) D^0)$ and   $N(\Lambda_b \to \Lambda(\to p \pi^-) D_{1,2})$  would be $2\times 10^4$ and $ 2 \times 10 ^{3}$,  respectively,   providing enough data points to reconstruct the full angular distributions,  and allowing the experiments to extract $\gamma$.

%There will be enough number of events in the future for our proposed observables to be observerd.

\section{Conclusion}\label{conclusion}
We have systematically studied the decays of ${\bf B}_b \to {\bf B}_nD$, and discussed the feasibilities of the experimental measurements. 
Remarkably, the process of $\Xi_b^- \to \Xi^- D$   contains only one  quark diagram, and therefore provides an ideal place to extract the weak phase. 
We have shown that the Wolfenstein parameters of the CKM matrix  can be extracted by $ \overline{\rho} = (R_1 - R_2)/2 $ and $ \overline{\eta} =  \sqrt{ 	R_1+R_2  - \frac{1}{4} ( R_1  - R_2 )^2	-1}$, which are feasible to be measured at LHC Run3.

 As the baryons can be polarized, we have demonstrated  that the decays of the b-baryons provide  much richer observables compared to the mesons. 
All the possible observables have been parameterized with 9 real parameters, which allows the future experiments to extract the CP violating unitarity angle of  $\gamma$
in the CKM matrix. 
At LHCb Run4 , a complete study of $ \Lambda_b \to \Lambda(\to p \pi^-) D$  on the angular distribution has shown to be promising.

On the other hand, to get quantitative results, the decay observables have been  studied with the factorization ansatz. In particular, we have found that  ${\cal B}( \Xi_b^- \to \Xi^- D) = ( 9.7 \pm 1.6, 1.4 \pm 0.3 , 6.9 \pm 1.2 , 4.2 \pm 1.7) \times 10^{-6}$ for $D = D^0,\overline{D}^ 0 , D_1 ,D_2$, respectively. We have also 
estimated
 the numerical results of $\Lambda_b \to \Lambda D$, which are  compatible with those in the  literature.

\begin{acknowledgments}
	This work is supported in part by the National Key Research and Development Program of China under Grant No. 2020YFC2201501 and  the National Natural Science Foundation of China under Grant No. 12147103.
\end{acknowledgments}

\appendix

	\section{Angular analysis and helicity formlism}
In this appendix, we would like to compare the pros and cons between the traditional approach and the helicity formalism. First, we briefly review the traditional approach toward the angular distribution~\cite{Pakvasa:1990if}.

	According to the Lorentz structure,
 the amplitudes of ${\bf B}_b \to {\bf B}_n D$ are traditionally parameterized as 
	\begin{equation}\label{equ}
M = i \overline{u}\left(  A - B \gamma_5 \right)  {u}_b \,,
	\end{equation}
where $A$ and $B$ are parameters to be determined by theories or experiments. 
Notice that with  Eq.~\eqref{equ}, we have chosen the spinor representations for ${\bf B}_{b,n}$.  Traditionally, Eq.~\eqref{equ} is recasted as
	\begin{equation}\label{key1}
		M=\chi_ {n}^{\dagger}(S+P \sigma \cdot \mathbf{q}) \chi_{b},
	\end{equation}
with 
\begin{equation}
S = i \sqrt{ 2M_b ( E_n+ M_n)} A \,,~~~~~P =-  i \sqrt{ 2M_b ( E_n-  M_n)}B
\end{equation}
	where $\chi_{b(n)}$ is the two-component spinor of ${\bf B}_{b(n)}$,  and  $ \mathbf{q} $ is the unit vector of the  ${\bf B}_n$'s   3-momentum  at the ${\bf B}_b$ rest frame.
	The symbols of $``S''$ and $``P''$ are related to  $L=0$ and $L=1$, respectively, where $L$ stands for the orbital angular momentum. 
	However, it is well known that the orbital angular momentum is ill-defined for massless particles, and thus incompatible to the special relativity. 

	By squaring the amplitudes and noting 
	\begin{equation}
\chi_b \chi_b ^ \dagger = \left(
\frac{1}{2} + \frac{1}{2} {\bf  s}_{b(n)}\cdot  {\sigma}
\right)\,,
	\end{equation}
we arrive 	
	\begin{equation}\label{key}
		\begin{aligned}
 1+\alpha_j \mathbf{q} \cdot(\mathbf{s}_b+\mathbf{s}_n)+(\mathbf{s}_b \cdot \mathbf{q})(\mathbf{s}_n \cdot \mathbf{q}) 
		+\beta_j  \mathbf{s}_b \times \mathbf{q} \cdot \mathbf{s}_n+\gamma_j \mathbf{q} \times(\mathbf{s}_b \times \mathbf{q}) \cdot \mathbf{s}_n  ,
		\end{aligned}
	\end{equation}
where ${\bf s}_{b(n) }$ satisfies
\begin{equation}
\left( {\bf s}_{b(n) } \cdot \sigma \right)\chi_{b(n)}  = \chi_{b(n)} \,. 
\end{equation}
Naively, one would tend to interpret ${\bf s}_{b(n)}$ as the spin of ${\bf B}_{b(n)}$. Nonetheless,
if we adopt such interpretation, Eq.~\eqref{key} would force us to commit  $({\bf s}_{b,n} )_x$, 
$({\bf s}_{b,n} )_y$ and $({\bf s}_{b,n} )_z$ commute with each others, as the notation suggests that ${\bf B}_{b,n}$ are the eigenstates of them simultaneously. 
Although the outcomes   might be correct,  the interpretation is fatally wrong on the theoretical aspect~\cite{TV,sizable}. 
Moreover, in the experiments, spins  can not be measured directly, and it is hard to deduce physical observables from  Eq.~\eqref{key}. 

In comparison,
the helicity formalism is outstanding on many aspects. 
It is perfectly compatible with  massless objects, and
 the angular distributions of sequential decays  can be easily deduced. 
For two-body systems, the eigenstates of helicities and angular momenta are constructed as
\begin{eqnarray}\label{HelicityStates}
	|\lambda_1\,, \lambda_2 , J, J_z \rangle = 
	\frac{1}{(2J+1 )\pi}
	\int d\cos \theta d \phi  |p, \theta , \phi , \lambda_1\,, \lambda_2\rangle  e^{iJ_z\phi} d^J(\theta) ^{J_z}\,_{\lambda_1 - \lambda_2}\,,
\end{eqnarray}
with 
\begin{equation}
| p,\theta, \phi,\lambda_1,\lambda_2  \rangle  = R_z(\phi) R_y(\theta ) \left( | f_1; \vec{p}= p \vec{z} , \lambda_1\rangle \otimes 
| f_2; \vec{p}= - p \vec{z} , \lambda_2\rangle \right) \,,
\end{equation}
where $f_{1,2}$ stand for the first and the second particles with opposite 3-momenta, $\lambda_{1,2}$ are the helicities of $f_{1,2}$, $J$ and $J_z$ are the angular momentum and its $z$ component of the systems, respectively, $d$ stands for  the Wigner-$d$ matrix,  and $R_{y,z}$ are the rotation operators pointing toward  $\hat{y}(\hat{z})$. 

For the weak decays of $i \to f_1 f_2$, where $i$ is an arbitrary particle pointing toward $\hat{z}$, the angular distributions are given  as 
\begin{equation}
\frac{\partial^2 \Gamma }{\partial \phi \partial \cos\theta } \propto \sum_{\lambda_1,\lambda_2} \left| 
\langle p, \theta , \phi ,\lambda_1 ,\lambda_2 | {\cal H}_{eff}| i; J , J_z\rangle
\right| ^2
\,. 
\end{equation}
The helicities of the outgoing particles must be summed over as they are  not distinguishable in the experiments.
By inserting the identity
\begin{equation}\label{B3}
	1 = \sum_{J,J_z,\lambda_1,\lambda_2}  \frac{4\pi }{2 J+1 }| J , J_z, \lambda_1 , \lambda_2 \rangle \langle  J , J_z, \lambda_1 , \lambda_2 |\,,
\end{equation}
we arrive  at 
\begin{equation}\label{B5}
	\frac{\partial^2 \Gamma }{\partial \phi \partial \cos \theta} \propto \sum_{\lambda_1,\lambda_2}  \left|
	e^{iJ_z\phi} d^{J}(\theta) ^{J_z} \,_{\lambda_1 - \lambda_2}  H^{\lambda_1\lambda_2} 
	\right|^2 \,,
\end{equation}
with 
\begin{equation} \label{HeliDef}
H^{\lambda_1,\lambda_2 } \equiv \langle J, J_z,\lambda_1,\lambda_2|{\cal H}_{eff} | i; J , J_z\rangle \,.
\end{equation}
Here, $H^{\lambda_1,\lambda_2}$ can not depend on $J_z$ as  ${\cal H}_{eff}$ is a scalar, and the exponential in Eq.~\eqref{B5} can clearly be omitted. 
We can see that the amplitudes are now decomposed into two parts; the kinematic part is described by the Wigner-$d$ matrices, while the dynamical one
by $H^{\lambda_1,\lambda_2}$.  
%{\color{red} 
Using the inverse of Eq.~\eqref{HelicityStates}, given as
\begin{equation}
| p \hat{z}, \lambda_1 ,\lambda_2 \rangle = \sum_{J} | J,J_z = \lambda_1-\lambda_2,\lambda_1 ,\lambda_2 \rangle\,,
\end{equation}
we arrive at 
\begin{equation}\label{newadded}
H^{\lambda_1,\lambda_2}  = \langle p \hat{z}, \lambda_1 ,\lambda_2 | {\cal H}_{eff} | i; J , \lambda_1 - \lambda_2\rangle\,,
\end{equation}
which is handy in computing the numerical results.
%}

Angular distributions of sequential decays can be obtained by applying the above method multiple times. 
For  ${\bf B}_b \to {\bf B}_n(\to {\bf B}_n' \pi )D_j$,  we have 
\begin{equation}
{\cal D}_j \propto \sum_{\lambda_n', J_z} \rho_{J_z J_z}  \left|\sum_{\lambda_n}
H^{ \lambda_n }_j H'^{\lambda_n'} e^{ i \lambda_n  \phi} d^{\frac{1}{2}} (\theta)^{J_z}\,_{\lambda_n} d^{\frac{1}{2}} (\theta_1)^{\lambda_n}\,_{\lambda_n' } 
\right|\,,
\end{equation}
where $H_j^{\lambda_n}$ describes the dynamic of  ${\bf B}_b \to {\bf B}_n D_j$  and $H^{\lambda_n'}$  of ${\bf B}_n \to {\bf B}_n' \pi$. Here, $\rho$ is the polarization density matrix of ${\bf B}_b$, given as 
\begin{equation}
\rho = \left( 
\begin{array}{cc}
\frac{1}{2}\left(1 + P_b\right)  & 0\\
0 &  \frac{1}{2}\left(1 -  P_b\right) 
\end{array}
\right) \,.
\end{equation}
The great advantage of  the helicity formalism is that we do not need to write down the explicit representations of the particles for obtaining the angular distributions. The whole analysis bases only on the group theory. 

\section{Estimations of  the numbers of the produced baryons }
In this appendix, we estimate the numbers of the events that can be reconstructed by the  experiments. 
%The production rate and the mass of $\Xi_b^-$ are measured by  $\Xi_b^- \to \Xi^- J/\psi  $ at LHCb. 
The  production ratios  at LHCb Run1 and Run2  are reported as~\cite{LHCb:2019sxa}
\begin{equation}
	\frac{  	f_{\Xi_b^-}  }{f_{\Lambda_b} } = (6.7 \pm 2.1 , 8.2 \pm 2.7 ) \times 10 ^{-2} \,, 
\end{equation}
respectively, where $f_{{\bf  B}_b} $ are the production rates of ${\bf B}_b$ .
At LHCb Run1 and Run2,
taking $\mathcal{B}(\Lambda_b \to \Lambda J/\psi)=(5.8\pm 0.8 )\times 10^{-4}$~\cite{pdg, Gutsche:2013oea} and  
$  	N(\Lambda_b \to \Lambda J/\psi )=(1.33,1.48)\times 10^{4}$~\cite{LHCb:2019sxa}, 
one finds that
%$ {\cal N}_{\Lambda_b} ({\cal N}_{\Xi_b^-} )$  are found to be  
\begin{equation}
	{\cal N}_{\Lambda_b} ({\cal N}_{\Xi_b^-} )=5.80\times 10^{9}~(3.89\times10^{8}),
\end{equation}  
%at  LHCb Run1,  
and  
\begin{equation}
	{\cal N}_{\Lambda_b} ({\cal N}_{\Xi_b^-} )=5.86\times 10^{9}~(4.81\times10^{8}),
\end{equation}
% at LHCb Run2. 
respectively.
On the other hand,  at LHCb Run3 and  Run4,
we have  \begin{equation}
	{\cal N}_{\Lambda_b}({\cal N}_{\Xi_b^-} )=1.10\times 10^{11}~(9.01\times10^{9}),
\end{equation}  
and
\begin{equation}
	{\cal N}_{\Lambda_b}({\cal N}_{\Xi_b^-} )= 1.83\times 10^{11}~(1.50\times10^{10}),
\end{equation}respectively.
Here, we have used that
the integrated luminosity  of  
LHCb Run3~(4)  is  18.75~(31.25) times larger  than that of  LHCb Run2~\cite{CERN:2017HLLHC}.


\begin{thebibliography}{9}




%\bibitem{LHCb:2020kho}
%[LHCb],
%LHCb-CONF-2020-003.
\bibitem{LHCb:2017hkl}
R.~Aaij \textit{et al.} [LHCb],
JHEP \textbf{03}, 059 (2018); 
R.~Aaij \textit{et al.} [LHCb],
JHEP \textbf{06}, 084 (2018).


\bibitem{Gronau:1990ra}
M.~Gronau and D.~London,
Phys. Lett. B \textbf{253}, 483 (1991);
M.~Gronau and D.~Wyler,
Phys. Lett. B \textbf{265}, 172 (1991).

\bibitem{Atwood:1996ci}
D.~Atwood, I.~Dunietz and A.~Soni,
Phys. Rev. Lett. \textbf{78}, 3257 (1997);
D.~Atwood, I.~Dunietz and A.~Soni,
Phys. Rev. D \textbf{63}, 036005 (2001).

\bibitem{Grossman:2002aq}
Y.~Grossman, Z.~Ligeti and A.~Soffer,
Phys. Rev. D \textbf{67}, 071301 (2003); 
A.~Giri, Y.~Grossman, A.~Soffer and J.~Zupan,
Phys. Rev. D \textbf{68}, 054018 (2003);
A.~Bondar and A.~Poluektov,
Eur. Phys. J. C \textbf{47}, 347 (2006).

\bibitem{CKM} 
CKMfitter group, J. Charles {\it et al.}, 
%Current status of the standard model CKM fit and constraints on ∆F = 2 new physics, 
Phys. Rev. D {\bf 91} 073007 (2015). 
% ,
% updated results and plots available at http://ckmfitter.in2p3.fr/.

\bibitem{UT}
UTfit collaboration, M. Bona {\it et al.}, 
%The unitarity triangle fit in the standard model and hadronic parameters from lattice QCD: A reappraisal after the measurements of ∆ms and BR(B → τ ντ ), 
JHEP {\bf 10},  081 (2006).




\bibitem{LBCP}
R.~Aaij \textit{et al.} [LHCb],
%``Searches for $\Lambda^0_{b}$ and $\Xi^{0}_{b}$ decays to $K^0_{\rm S} p \pi^{-}$ and $K^0_{\rm S}p K^{-}$ final states with first observation of the $\Lambda^0_{b} \rightarrow K^0_{\rm S}p \pi^{-}$ decay,''
JHEP \textbf{04}, 087 (2014); 
R.~Aaij \textit{et al.} [LHCb],
%``Observations of $\Lambda_b^0 \to \Lambda K^+\pi^-$ and $\Lambda_b^0 \to \Lambda K^+K^-$ decays and searches for other $\Lambda_b^0$ and $\Xi_b^0$ decays to $\Lambda h^+h^{\prime -}$ final states,''
JHEP \textbf{05}, 081 (2016);
R.~Aaij \textit{et al.} [LHCb],
%``Measurements of $CP$ asymmetries in charmless four-body $\Lambda_b^0$ and $\Xi_b^0$ decays,''
Eur. Phys. J. C \textbf{79}, 745 (2019). 


\bibitem{LHCb:2022piu}
R.~Aaij \textit{et al.} [LHCb],
%``Observation of the decay $ \Lambda_b^0\rightarrow \Lambda_c^+\tau^-\overline{\nu}_{\tau}$,''
arXiv:2201.03497 [hep-ex].


\bibitem{Lambdamu}
R.~Aaij \textit{et al.} [LHCb],
%``Differential branching fraction and angular analysis of $\Lambda^{0}_{b} \rightarrow \Lambda \mu^+\mu^-$ decays,''
JHEP \textbf{06}, 115 (2015).

\bibitem{LHCb:2021byf}
R.~Aaij \textit{et al.} [LHCb],
%``Measurement of the photon polarization in $\Lambda_{b}^{0}$ $\to$ $\Lambda$ $\gamma$ decays,''
Phys. Rev. D \textbf{105},  L051104 (2022). 


\bibitem{Jpsi}
G.~Aad \textit{et al.} [ATLAS],
%``Measurement of the parity-violating asymmetry parameter $\alpha_b$ and the helicity amplitudes for the decay $\Lambda_b^0\to J/\psi+\Lambda^0$ with the ATLAS detector,''
Phys. Rev. D \textbf{89},  092009 (2014);
A.~M.~Sirunyan \textit{et al.} [CMS],
%``Measurement of the $\Lambda_b$ polarization and angular parameters in $\Lambda_b\to J/\psi\, \Lambda$ decays from pp collisions at $\sqrt{s}=$ 7 and 8 TeV,''
Phys. Rev. D \textbf{97},  072010 (2018);
R.~Aaij \textit{et al.} [LHCb],
%``Measurement of the $\Lambda^0_b\rightarrow J/\psi\Lambda$ angular distribution and the $\Lambda^0_b$ polarisation in $pp$ collisions,''
JHEP \textbf{06}, 110 (2020).

\bibitem{Observables}
M.~He, X.~G.~He and G.~N.~Li,
%``CP-Violating Polarization Asymmetry in Charmless Two-Body Decays of Beauty Baryons,''
Phys. Rev. D \textbf{92}, 036010 (2015).
\bibitem{TV}
C.~Q.~Geng and C.~W.~Liu,
%``Time-reversal asymmetries and angular distributions in \ensuremath{\Lambda}$_{b}$ \textrightarrow{} \ensuremath{\Lambda}V,''
JHEP \textbf{11}, 104 (2021). 

\bibitem{sizable}
C.~W.~Liu and C.~Q.~Geng,
%``Sizable time-reversal violating effects in bottom baryon decays,''
JHEP \textbf{01}, 128 (2022).



\bibitem{Theor}
C.~D.~Lu, Y.~M.~Wang, H.~Zou, A.~Ali and G.~Kramer,
%``Anatomy of the pQCD Approach to the Baryonic Decays Lambda(b) ---\ensuremath{>} p pi, p K,''
Phys. Rev. D \textbf{80}, 034011 (2009);
Y.~K.~Hsiao and C.~Q.~Geng,
%``Direct CP violation in $\Lambda_b$ decays,''
Phys. Rev. D \textbf{91}, 116007 (2015);
Y.~M.~Wang and Y.~L.~Shen,
%``Perturbative Corrections to $\Lambda_b \to \Lambda$ Form Factors from QCD Light-Cone Sum Rules,''
JHEP \textbf{02}, 179 (2016);
Z.~X.~Zhao,
%``Weak decays of heavy baryons in the light-front approach,''
Chin. Phys. C \textbf{42}, 093101 (2018);
J.~Zhu, Z.~T.~Wei and H.~W.~Ke,
%``Semileptonic and nonleptonic weak decays of $\Lambda_b^0$,''
Phys. Rev. D \textbf{99}, 054020 (2019);
Y.~S.~Li and X.~Liu,
%``Restudy of the color-allowed two-body nonleptonic decays of bottom baryons \ensuremath{\Xi}b and \ensuremath{\Omega}b supported by hadron spectroscopy,''
Phys. Rev. D \textbf{105},  013003 (2022); 
C.~Q.~Zhang, J.~M.~Li, M.~K.~Jia and Z.~Rui,
%``Nonleptonic two-body decays of $\Lambda_b\rightarrow \Lambda_c \pi, \Lambda_cK$ in the perturbative QCD approach,''
arXiv:2202.09181;
J.~J.~Han, Y.~Li, H.~n.~Li, Y.~L.~Shen, Z.~J.~Xiao and F.~S.~Yu,
%``$\Lambda_b\to p$ transition form factors in perturbative QCD,''
arXiv:2202.04804;
C.~Q.~Geng, C.~W.~Liu, Z.~Y.~Wei and J.~Zhang,
%``Weak Radiative Decays of Anti-triplet Bottomed Baryons in Light-Front Quark Model,''
Phys. Rev. D  \textbf{05},  073007 (2022).


\bibitem{Zhu:2018jet}
J.~Zhu, Z.~T.~Wei and H.~W.~Ke,
%``Semileptonic and nonleptonic weak decays of $\Lambda_b^0$,''
Phys. Rev. D \textbf{99},  054020 (2019).


\bibitem{Geng:2020ofy}
C.~Q.~Geng, C.~W.~Liu and T.~H.~Tsai,
%``Nonleptonic two-body weak decays of $\Lambda_b$ in modified MIT bag model,''
Phys. Rev. D \textbf{102},  034033 (2020);
C.~W.~Liu and C.~Q.~Geng,
%``Center of mass motion in bag model,''
arXiv:2205.08158.


\bibitem{Gutsche:2013oea}
T.~Gutsche, M.~A.~Ivanov, J.~G.~K\"orner, V.~E.~Lyubovitskij and P.~Santorelli,
%``Polarization effects in the cascade decay $Λ_b → Λ (→ pπ^-) + J/ψ (→ ℓ^+ℓ^-)$ in the covariant confined quark model,''
Phys. Rev. D \textbf{88},  114018 (2013); 
Z.~P.~Xing, F.~Huang and W.~Wang,
%``Angular distributions for $\Lambda_b\to \Lambda^*_J (pK^-)J/\psi$ Decays,''
arXiv:2203.13524.

\bibitem{Giri:2001ju}
A.~K.~Giri, R.~Mohanta and M.~P.~Khanna,
%``Possibility of extracting the weak phase gamma from Lambda(b) ---\ensuremath{>} Lambda D0 decays,''
Phys. Rev. D \textbf{65}, 073029 (2002).

\bibitem{Zhang:2021sit}
S.~Zhang, Y.~Jiang, Z.~Chen and W.~Qian,
%``Sensitivity studies on the CKM angle $\gamma$ in $\Lambda_b^0 \to D\Lambda$ decays,''
arXiv:2112.12954.

\bibitem{pdg}
P.~A.~Zyla \textit{et al.} [Particle Data Group],
%``Review of Particle Physics,''
PTEP \textbf{2020},  083C01 (2020).

\bibitem{Buras:1991jm}
A.~J.~Buras, M.~Jamin, M.~E.~Lautenbacher and P.~H.~Weisz,
%``Effective Hamiltonians for $\Delta S = 1$ and $\Delta B = 1$ nonleptonic decays beyond the leading logarithmic approximation,''
Nucl. Phys. B \textbf{370}, 69 (1992).
\bibitem{Zhang:2021yul}
W.~X.~Zhang, H.~Xu and D.~Jia,
%``Masses and magnetic moments of hadrons with one and two open heavy quarks: Heavy baryons and tetraquarks,''
Phys. Rev. D \textbf{104}, 114011 (2021).


\bibitem{BESIII:2020khq} 
M.~Ablikim \textit{et al.} [BESIII], %``Model-independent determination of the relative strong-phase difference between $D^0$ and $\bar{D}^0\rightarrow K^0_{S,L}\pi^+\pi^-$ and its impact on the measurement of the CKM angle $\gamma/\phi_3$,'' 
Phys. Rev. D \textbf{101}, 112002 (2020). 

\bibitem{LHCb:2019sxa}
R.~Aaij \textit{et al.} [LHCb],
%``Measurement of the mass and production rate of $\Xi_b^-$ baryons,''
Phys. Rev. D \textbf{99},  052006 (2019).




\bibitem{Pakvasa:1990if}
S.~Pakvasa, S.~P.~Rosen and S.~F.~Tuan,
%``Parity Violation and Flavor Selection Rules in Charmed Baryon Decays,''
Phys. Rev. D \textbf{42}, 3746 (1990).

\bibitem{CERN:2017HLLHC}
G.~Apollinari, I.~B\'ejar Alonso, O.~Br\"uning, P.~Fessia, M.~Lamont, L.~Rossi, L.~Tavian,
%``High-Luminisoty Large Hadron Collider(HL-LHC).Technical Design Report V.0.1,''
CERN Yellow Reports: Monographs, Vol.4/2017, CERN-2017-007-M.

\end{thebibliography}
\end{document}